# Large nonlinear Hall effect and Berry curvature in KTaO$_3$ based two-dimensional electron gas

Jinfeng Zhai[1†], Mattia Trama[2,3,4†], Hao Liu[1], Zhifei Zhu[1], Yinyan Zhu[1], Carmine Antonio Perroni[5,6], Roberta Citro[2,3,6*], Pan He[1,7,8*] and Jian Shen[1,7,8,9,10,11*]

[1]State Key Laboratory of Surface Physics and Institute for Nanoelectronic devices and Quantum computing, Fudan University, Shanghai 200433, China

[2]Physics Department "E.R. Caianiello", Universitá degli studi di Salerno, Via Giovanni Paolo II, 132, I-84084 Fisciano (Sa), Italy

[3]INFN - Sezione di Napoli, Compesso Univ. Monte S. Angelo, I-80126 Napoli, Italy

[4]Institute for Theoretical Solid State Physics, IFW Dresden, Helmholtzstr. 20, 01069 Dresden, Germany.

[5]Physics Department "Ettore Pancini", Universitá degli studi di Napoli "Federico II", Complesso Univ. Monte S. Angelo, Via Cintia, I-80126 Napoli, Italy

[6]CNR-SPIN Napoli Unit, Complesso Univ. Monte S. Angelo, Via Cintia, I-80126 Napoli, Italy

[7]Shanghai Qi Zhi Institute, Shanghai 200232, China

[8]Zhangjiang Fudan International Innovation Center, Fudan University, Shanghai 201210, China

[9]Department of Physics, Fudan University, Shanghai, China

[10]*Shanghai Research Center for Quantum Sciences, Shanghai, China*

[11]*Collaborative Innovation Center of Advanced Microstructures, Nanjing, China*

**ABSTRACT:** The two-dimensional electron gas (2DEG) at oxide interfaces exhibits various exotic properties stemming from interfacial inversion symmetry breaking. In this work, we report the emergence of large nonlinear Hall effects (NHE) in the $LaAlO_3/KTaO_3$(111) interface 2DEG under *zero* magnetic field. Skew scattering was identified as the dominant origin based on the cubic scaling of nonlinear Hall conductivity with longitudinal conductivity and the threefold symmetry. Moreover, a gate-tunable NHE with pronounced peak and dip was observed and reproduced by our theoretical calculation. These results indicate the presence of Berry curvature hotspots and thus a large Berry curvature triple at the oxide interface. Our theoretical calculations confirm the existence of large Berry curvatures from the avoided crossing of multiple 5*d*-orbit bands, orders of magnitude larger than that in transition-metal dichalcogenides. NHE offers a new pathway to probe the Berry curvature at oxide interfaces, and facilitates new applications in oxide nonlinear electronics.

**KEYWORDS:** oxide interface, two-dimensional electron gas, nonlinear Hall effect, gate-tunability, Berry curvature triple

Since the discovery of a highly mobile two-dimensional electron gas (2DEG) at LaAlO$_3$/SrTiO$_3$ interface[1], complex oxide heterostructures have attracted great interests. The oxide interface 2DEG exhibits exotic phenomena, including magnetism[2], superconductivity[3], and electric-field control of conductivity[4], which are absent in their parent compounds. Due to the interfacial inversion symmetry breaking, Rashba spin-orbit coupling (SOC) leads to spin-polarized electronic bands, which facilitates efficient spin-charge interconversions[5, 6] and novel spin-dependent magnetoresistances[7]. These findings have sparked the field of oxide spintronics[8]. Recently, much of the attention has shifted to KTaO$_3$ (KTO) based 2DEG heterostructures, such as LaTiO$_3$/KTO[9], LaAlO$_3$(LAO)/KTO[10], and EuO/KTO[11]. KTO is a band insulator, similar to SrTiO$_3$ with a high permittivity and quantum paraelectricity[12]. The atomic SOC, however, is much stronger in KTO (~0.4 eV) than that in SrTiO$_3$. The KTO-based 2DEG was found to show a higher critical temperature of superconductivity[13, 14], a larger spin-splitting of Fermi surface[15], and a more efficient spin-charge interconversion[16]. Due to the strong SOC, the absence of inversion symmetry at the interface, and multi-5$d$-orbits of KTO based 2DEGs, band anti-crossings are expected to generate an enhanced Berry curvature, which encodes the geometric properties of the electronic wavefunctions. However, demonstrating its existence and effect on electron transports have not been explored[17].

Linear Hall effect is forbidden under time-reversal symmetry[18]. However, the second-order nonlinear Hall effect (NHE) does not obey this constraint, but it requires inversion symmetry breaking. Following a predication by Sodemann and Fu[19], NHE was recently observed in non-centrosymmetric $WTe_2$ under time-reversal symmetric conditions[20, 21]. NHE arises either from the Berry curvature dipole (BCD)[19] or from the impurity scatterings, such as skew scattering[22, 23] and side jump[24]. While the BCD driven NHE was found in crystals with one mirror reflection at maximum[20, 25], the skew scattering originated NHE was recently observed in materials with three mirror reflections[22, 26]. Skew scattering is due to the inherent chirality of itinerant electrons and exhibits a close relation to the Berry curvature[22, 23, 26]. It is allowed in all noncentrosymmetric time-reversal invariant materials. So far, NHE has been widely studied in non-oxide materials[24]. Although oxide heterostructures represent a large family of materials with inversion symmetry breaking, the demonstration of NHE is still lacking in these materials. The mechanisms for generating Berry curvature include SOC, orbit hybridization, etc.[27]. It is thus highly desirable to study the Berry curvature related NHE in the KTO-based 2DEG, which exhibits a large SOC and multiple *d*-orbit hybridizations.

In this paper, we report the observation of NHE in the 2DEG at LAO/KTO(111) interface up to room temperature (RT) under *zero* magnetic field. The nonlinear Hall conductivity was found to decrease dramatically with rising temperature. Its scaling law with the longitudinal conductivity and the threefold symmetry establish skew scattering as the dominant physical origin. Moreover, a large and highly tunable NHE with peak and dip structures was observed by changing back-gate voltages, indicating the existence of large Berry curvature triples with sign changes. Our theoretical calculations show that the interplay of quantum confinement, strong SOC, and multi-*d*-orbits results in the generation of Berry curvature hotspots at the anti-crossing bands and distributions of Berry curvature with sign changes. NHE provides a new way to study topologically non-trivial band structures at the correlated oxide interfaces that differs from standard magneto-transport measurements[28]. The emergence of NHE will inject new momentum to the study of oxide heterostructures.

The LAO/KTO interfacial 2DEGs were prepared by growing a LAO overlayer on single-crystalline KTO(111) substrates using pulsed laser deposition (see Supporting Information S1). The lattice structure of the KTO(111) surface is schematically shown in Figure 1a. Previous studies have shown that oxygen vacancies near the LAO/KTO interface can effectively electron dope the interfacial KTO[29]. Hall bar devices (Figure 1b) were fabricated via photolithography for transport measurements. As shown in Figure 1c, the linear and nonlinear electric transports can be separated by respectively measuring the first harmonic ($V_i^{1\omega}$) and the second harmonic voltages ($V_i^{2\omega}$) under an ac current $I_x = I\sin\omega t$ using the lock-in techniques (Supporting Information S1) with *i* representing the current direction *x* or the transverse direction *y*. The LAO/KTO interface exhibits metallic behavior, as the sheet resistance $\rho_{sheet}$ increases monotonically with rising temperature in Figure 1d. Correspondingly, the carrier mobility μ shows a dramatic decrease with rising temperature in Figure 1e.

To explore the NHE under time-reversal symmetry, we measured the second-harmonic Hall voltage $V_y^{2\omega}$ at zero magnetic field. Figure 2a displays a sizable $V_y^{2\omega}$ for the current applied along the $[\bar{1}, 1, 0]$ direction with a quadratic current dependence, i.e., $V_y^{2\omega} \propto I^2$. In contrast, a much smaller $V_y^{2\omega}$ was observed for the current applied along the $[\bar{1}, \bar{1}, 2]$ direction (Supporting Information S2a). For both current directions, $V_y^{2\omega}$ changes sign when reversing the current and the corresponding Hall probes simultaneously (schematically shown in the inset of Figure 2a). We further confirmed that $V_y^{2\omega}$ is independent of the ac frequency used and has a negligible contribution from the contact misalignment of Hall device (Supporting Information S3). The dc biased ac measurements were also performed to further verify the observation of NHE (Supporting Information S3). These results unambiguously demonstrate the existence of NHE at the LAO/KTO(111) interface under time-reversal symmetry.

To investigate the temperature dependence of NHE, the $V_x^{1\omega}(I)$ and $V_y^{2\omega}(I)$ curves were measured simultaneously at different temperatures in Figure 2b, and 2c, respectively. The linear dependence of $V_x^{1\omega}$ on $I$ indicates good ohmic contacts. The quadratic dependence of $V_y^{2\omega}$ on $I$ was observed at all the temperatures. The nonlinear Hall conductivity $\sigma_{yxx}^{(2)}$ can then be calculated using the formula $\sigma_{yxx}^{(2)} = -\frac{\sigma_{xx} V_y^{2\omega} L^2}{(V_x^{1\omega})^2 W}$, where $\sigma_{xx}$ is the linear longitudinal conductivity, $L$ ($W$) is the length (width) of the Hall bar device[26]. A dramatic decrease of $\sigma_{yxx}^{(2)}$ with rising temperature was found in Figure 2d. To understand such temperature dependence, the $\sigma_{yxx}^{(2)}/\sigma_{xx}$ is plotted as a function of $(\sigma_{xx})^2$ in Figure 3a. A good linear fit is shown as the red line. This scaling analysis can separate different contributions to the NHE. Specifically, the linear slope corresponds to a $(\sigma_{xx})^3$ contribution to $\sigma_{yxx}^{(2)}$, which represents the skew scattering driven NHE[21, 22]. The nonzero intercept of the fitting corresponds to a $\sigma_{xx}$ linear contribution to $\sigma_{yxx}^{(2)}$, which originates from BCD or side jump[19, 21]. Since KTO remains in the cubic phase down to 4.2 K[30], the persistence of a threefold rotational symmetry for the (111) surface excludes the existence of BCD in the LAO/KTO(111) 2DEG[15, 19]. Thus, the $\sigma_{xx}$ linear term is likely from the side jump, rather than BCD. This is different from SrTiO$_3$ based 2DEG that shows a cubic-to-tetragonal structural transition at 110 K, thus breaking the three-fold rotational symmetry along the [111] direction and enabling a finite BCD[31]. Moreover, a threefold angular dependence of NHE observed in Figure 3c. It further supports the skew scattering origin and excludes the BCD origin, as the two mechanisms give rise to totally different angular dependences[21, 22]. From the obtained fitting parameters, different contributions to $\sigma_{yxx}^{(2)}$ can be plotted as a function of $\sigma_{xx}$ in Figure 3b. We found NHE arises dominantly from skew scattering in the highly conductive regime[23, 26]. The wave packet of Bloch electron self-rotates with a finite Berry curvature, and its chirality reverses with changing the sign of Berry curvature as schematically shown in Fig. 3d[22]. Skew scattering arises from the chiral Bloch electron wave, whereby disorder deflects its motion in a preferred direction. It leads to the NHE in non-centrosymmetric materials as demonstrated previously in a gapped graphene and Bi$_2$Se$_3$[22, 23, 26]. Thus, the skew scattering related NHE acts as a hallmark of Berry curvature in the KTO based 2DEGs.

It is worth noting that a sizable $V_y^{2\omega}$ was observed even at room temperature (RT) under a moderate current (Supporting Information S2b). This was achieved with the help of a high sheet resistance at RT (6 kΩ per square)[32]. The voltage responsivity $\mathcal{R}_V$, defined as the ratio of the generated DC voltage to the power dissipation, is a figure of merit to quantify the conversion efficiency of rectifiers. $\mathcal{R}_V$ can be estimated as $\mathcal{R}_V = \frac{V_y^{2\omega}L}{\sigma_{xx}(V_x^{1\omega})^2 W} = 0.2$ V/W at RT and decreases with lowering temperature, differing from $\sigma_{yxx}^{(2)}$ (Supporting Information S4). To enhance $\mathcal{R}_V$ for future applications, the interface engineering has to be optimized and a large $\sigma_{yxx}^{(2)}$ and a small $\sigma_{xx}$ are highly desirable. We highlight the scalable sample preparation at RT makes KTO based 2DEGs promising materials for nonlinear device applications.

KTO has an extremely high dielectric constant (~5000 at 2K), making it an ideal oxide for field effect application[33]. To study the field effect on NHE, back-gate voltages $V_g$ were applied between the bottom of KTO substrate and the interfacial 2DEG (Figure 1c). Remarkably, while the longitudinal sheet resistance shows a monotonical decrease by two times with increasing $V_g$ from -40 V to +40 V in Figure 4a[10, 34], the nonlinear Hall resistance $R_{yx}^{2\omega} = V_y^{2\omega}/I$ shows a strong $V_g$ dependence with sign reversals, a peak and a dip in Figure 4b. The monotonical decrease of $\rho_{sheet}$ with $V_g$ can be mainly due to reason that more electron are occupying the bands, as estimated from the linear Hall effect measurements while scanning a magnetic field[14]. The variation of $V_g$ induces a $n_{2D}$ change of ~2.3×10$^{13}$ cm$^{-2}$ in Figure 4c, which corresponds to a sizable Fermi energy variation. As a Fermi energy sensitive property, $\sigma_{yxx}^{(2)}$ is plotted as a function of $V_g$ in Figure 4d with a nonmonotonic variation. Similar gate dependent results were obtained in other samples (Supporting Information S5). We note that such a large gate tunability of NHE has only been demonstrated before in 2D van der Waals materials[20, 26]. $\sigma_{yxx}^{(2)}$ obtained in LAO/KTO can be several times larger than that in WTe$_2$ [20, 21].

To gain more insight into the gate dependence, the electronic band structure of 2DEG on KTO (111) surface has been obtained in Figure 5a-c by performing a tight-binding supercell calculation (Supporting Information S6), which reproduces the angle-resolved photoemission spectroscopy results[15]. The 2DEG inherits the distinct properties of its parent bulk KTO, in addition to the quantum confinement effect at the surface. Specifically, the 2DEG is formed by the bands derived from the bulk $J = 3/2$ doublet, which further generates the high-order sub-bands under the spatial confinement along the KTO [111] direction[15]. Tight-binding calculations suggested a strong mixing of $d_{xy}$, $d_{xz}$, and $d_{yz}$ orbitals driven by SOC[15, 35] and a sixfold symmetric Fermi surface consisting of star-shaped contours and hexagonal contours in Figure 5c. Note the band structures for 2DEG residing on KTO (111) surface are different from that on the well-studied (001) surface[10, 34].

While the Rashba SOC has been widely studied at oxide interfaces[34], the Berry curvature generation has rarely been explored. In this work, we calculated the Berry curvature distribution $\Omega_z(\mathbf{k})$ for the 2DEG on KTO(111) surface (Supporting Information S6). The results for two representative bands are shown in Fig. 5d,e. The Berry curvature displays a significant momentum dependence and approaches maximum values (hotspots) at the $\mathbf{k}$ points in the vicinity of avoided band crossings, as the closer in energy the bands are, the more they contribute to the Berry curvature. The avoided crossing along the Γ-K direction extends over the full bandwidth of the system with a large Berry curvature as shown in Figure 5d, whereas the Berry curvature vanishes along Γ-M. The Berry curvature hotspots were also found close to the Γ-M direction when the different sub-bands intersect each other in Figure 5e. The distribution of Berry curvature is in line with the threefold symmetry of the KTO(111) surface, which enables a Berry curvature triple. It explains the threefold angular dependence of NHE observed in Figure 3c, showing a maximum value along $[\bar{1}, 1, 0]$ and a small one along $[\bar{1}, \bar{1}, 2]$ (see also Supporting Information S2). The typical Berry curvature near the anti-crossing band structures is around 2000 $Å^2$, which is about two order of magnitude larger than that in the well-studied transition-metal dichalcogenides (TMDC) at the $K$ points, such as $MoS_2$[36] ($\Omega_z \sim 9.88$ Å$^2$) and $WSe_2$[37] ($\Omega_z \sim 3.0$ Å$^2$). Another interesting observation is the presence of critical Fermi levels where Berry curvature changes sign from a positive hotspot to a negative one. The nonlinear Hall conductivity has been calculated in Figure 5f based on the Berry curvature distribution and skew scattering theory, whereby $\sigma_{yxx}^{(2)} \propto \frac{e^3 \tau^3}{p_F \tilde{\tau}}$ ($\tau$ is the symmetric scattering time, $\tilde{\tau}$ is the skew scattering time, $e$ is the electric charge and $p_F$ is the Fermi momentum)[22, 23]. The calculation results are in line with our observation of gate-dependent NHE with the peak and dip structures and sign changes in-between, reinforces the explanation by skew scattering. The skew scattering driven NHE is enhanced at the Fermi energy with Berry curvature hotspots[22, 26], as the skew scattering rate $\tilde{\tau}^{-1}$ is proportional to the Berry curvature triple $T(\epsilon_F) = 2\pi\hbar \int \frac{d^2k}{(2\pi)^2} \delta(\epsilon_F - \epsilon_k) \Omega_z(\bm{k}) \cos 3\theta_k$ in the threefold system[22]. $T(\epsilon_F)$ quantifies the strength of Berry curvature distribution on the Fermi surface. NHE shows peaks at the band filling to the anti-crossing of band structure. In this sense, we can map gate-dependent NHE to the topological band structure of oxide 2DEG[5]. To establish the correspondence, the theoretical carrier density versus Fermi level has been calculated in Supporting Information S7. A small discrepancy between the theoretical carrier density and the experimental one was observed to achieve the NHE peak. It can be due to the simplified model for the theoretical calculation, as reported in other studies[34].

The NHE was further observed in samples with different carrier densities, and it shows a similar temperature dependence (Supporting Information S8). The higher mobility sample shows a larger $\sigma_{yxx}^{(2)}$, consistent with skew scattering theory. In addition, we deposited other overlayer material on KTO(111) surface to induce the interface 2DEG, such as $Al_2O_3$. Similar NHE results have been obtained (Supporting Information S9). We may conclude that the NHE is a universal property for the KTO based 2DEGs, irrespective of its carrier density or overlayer materials used. Moreover, we demonstrated the NHE in $SrTiO_3$ based 2DEGs with a similar temperature dependence and scaling behavior with the longitudinal conductivity (Supporting Information S10). Noting conducting oxide interfaces include a large pool of materials[38], NHE are anticipated to exist in other oxide interfaces. NHE can be further enhanced by interface-engineering of the artificial 2DEGs, which can produce higher mobility samples[39]. The large Berry curvature in the KTO based 2DEGs is expected to induce a large intrinsic spin Hall effect, which waits for future investigations.

In conclusion, we demonstrated the emergence of NHE in the LAO/KTO(111) interface 2DEG at zero magnetic field, which sustains up to RT. A large nonlinear Hall conductivity was found, which can be a few times larger than that reported in $WTe_2$. With the absence of BCD and the existence of Berry curvature triple, skew scattering dominates the NHE in the LAO/KTO(111) 2DEG. The large SOC, quantum confinement, and multiorbital character have been utilized in this work to generate topological nontrivial bands with Berry curvature hotspots. The highly Fermi energy dependent Berry curvature distribution enables the nonlinear transport of Bloch electron with a large gate tunability. Our results will stimulate the search for topological non-trivial band structures in other oxide heterostructures from the perspective of NHE.

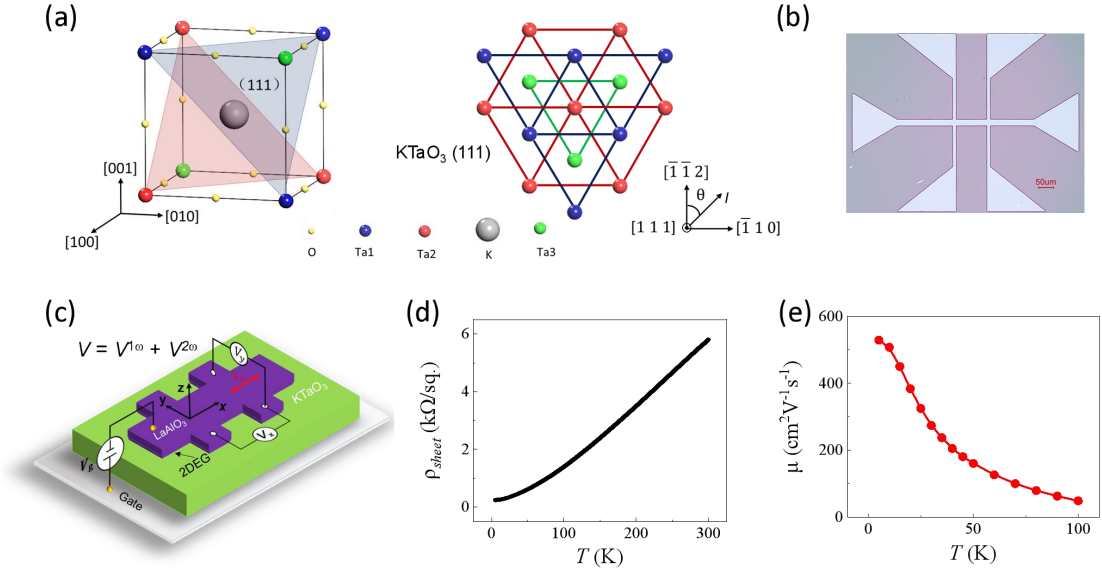

**Figure 1.** 2DEG at LaAlO$_3$/KTaO$_3$ (111) interface and its basic transport characterizations. (a) Schematic unit cell of the perovskite KTaO$_3$ with the (111) plane shaded (left panel). Top view of three consecutive Ta layers on the (111) surface (right panel). The $[\bar{1}\,1\,0]$ and $[\bar{1}\,\bar{1}\,2]$ axes are indicated. The current orientation angle θ with respect to the $[\bar{1},\bar{1},2]$ axis is indicated. (b) Optical image of a Hall bar. (c) Schematic of a field-effect device of LAO/KTO 2DEG and the electric harmonic measurements. A sinusoidal current $I^\omega$ was applied, and the first harmonic $V^{1\omega}$ and second harmonic $V^{2\omega}$ voltages were simultaneously measured at zero magnetic field along the longitudinal ($x$) and transverse ($y$) directions. The longitudinal voltage ($V_x$), transverse voltage ($V_y$) and gate voltage ($V_g$) are indicated. (d), (e) The sheet resistance $\rho_{sheet}$ (d) and mobility μ (e) as a function of temperature.

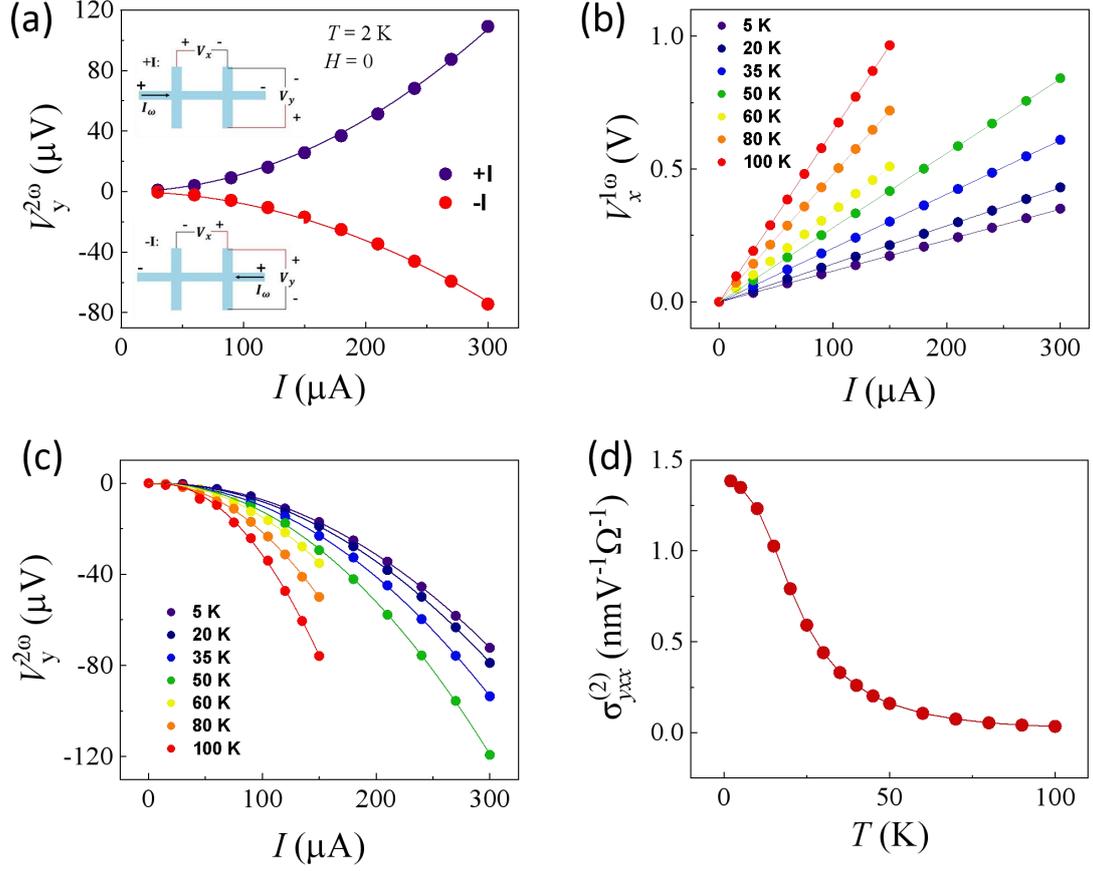

**Figure 2.** Observation of NHE and its temperature dependence in LAO/KTO(111) 2DEG. (a) The second-harmonic Hall voltage $V_y^{2\omega}$ versus ac amplitude $I$ for the current along $[\bar{1}10]$ axis. The solid curves are quadratic fits to the data. The inset show the schematic of current and voltage probe connections under opposite currents of $+I$ and $-I$. The arrow indicates the current direction during the first half period of sinusoidal current. The data were collected at $T = 2$ K. (b) The first-harmonic longitudinal voltage $V_x^{1\omega}$ as a function of $I$ at different temperatures. The solid lines are linear fits to the data. (c) The $V_y^{2\omega}$ versus $I$ at different temperatures. The solid curves are quadratic fits. (d) The nonlinear Hall conductivity $\sigma_{yxx}^{(2)}$ as a function of temperature. The data in (a-d) were measured under zero magnetic field in Sample 1.

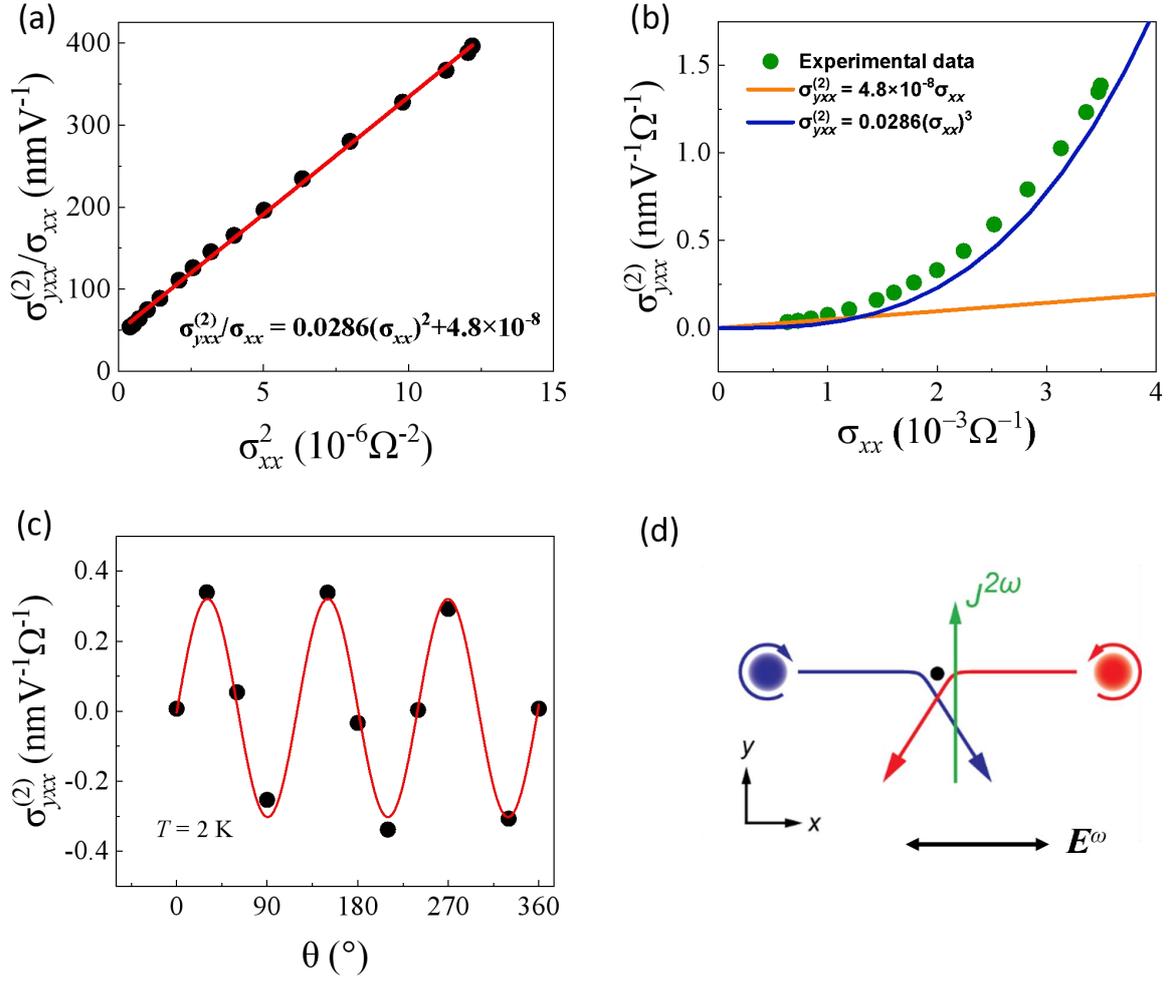

**Figure 3.** The scaling law of NHE and its three-fold symmetry. (a) The $\sigma^{(2)}_{yxx}/\sigma_{xx}$ versus $(\sigma_{xx})^2$. The red line is a linear fit to the experimental data. (b) The separated contributions to $\sigma^{(2)}_{yxx}$ as a function of $\sigma_{xx}$. The linear line (orange) represents the side jump contribution, while the cubic curve (blue) shows the skew scattering contribution. (c) The nonlinear Hall conductivity $\sigma^{(2)}_{yxx}$ versus the direction θ of current with respect to the $[\bar{1}, \bar{1}, 2]$ axis. The red line is a sin3θ fit to the experimental data. (d) Schematic plot of the skew scattering at LAO/KTO(111) interface and the induced nonlinear transverse transport. The blue and red balls represent self-rotating electron wave packets with opposite chirality (Berry curvature).

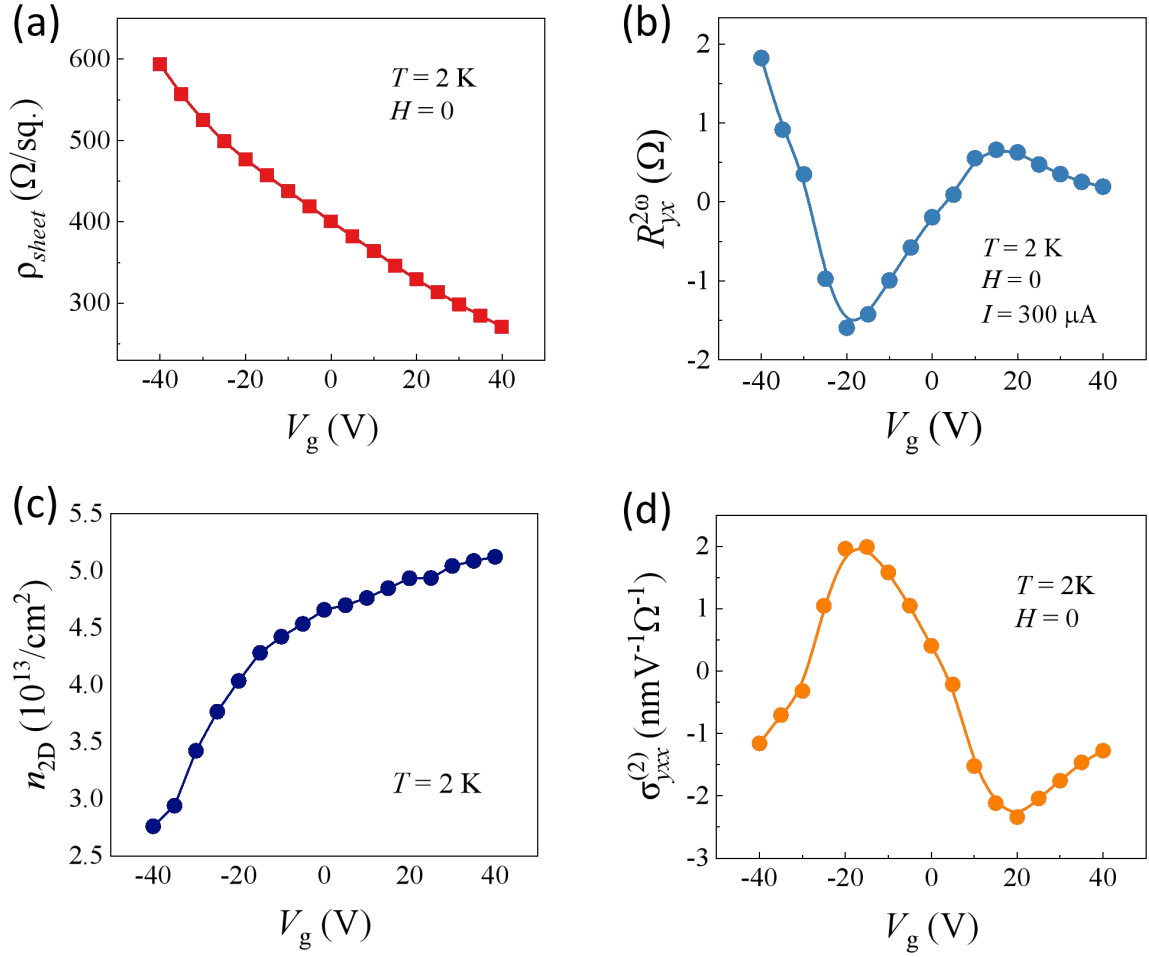

**Figure 4.** Back-gate dependence of NHE. (a) The sheet resistance ρ$_{sheet}$ versus gate voltage $V_g$. (b) The nonlinear Hall resistance $R_{yx}^{2\omega}$ versus $V_g$. (c) The carrier density $n_{2D}$ versus $V_g$. (d) $\sigma_{yxx}^{(2)}$ versus $V_g$. The data in (a-d) were measured under zero magnetic field and 2K in Sample 1.

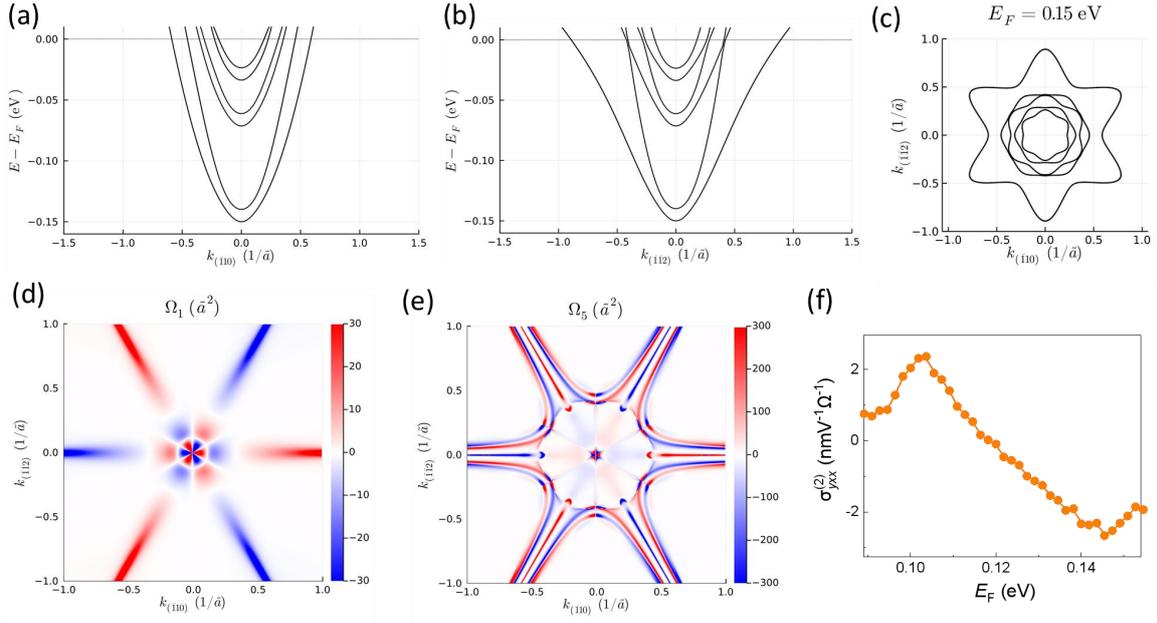

**Figure 5.** The calculated band structure and Berry curvature distribution for the 2DEG at KTO(111) surface. (a, b) Tight-binding supercell calculation of electronic band structures along the k $[\bar{1}10]$ (Γ-K) (a) and k $[\bar{1}\bar{1}2]$ (Γ-M) (b) directions for the 2DEG on KTO(111) surface. (c) Fermi surface for the benchmark choice of the Fermi level $E_F$ = 0.15 eV. (d-f) Berry curvature distribution in the k space for one of the first (d), and third (e) doublets of the electronic band structure. The color bar indicates the magnitude of Berry curvature with $\tilde{a} = 0.3988 \times \sqrt{2/3}$ nm. (f) Theoretical nonlinear Hall conductivity $\sigma_{yxx}^{(2)}$ calculated based on the Berry curvature triple $T$ using the skew scattering theory with the coefficients chosen to best fit the experimental results.

ASSOCIATED CONTENT

**Data Availability Statement**

All data needed to evaluate the conclusions in the paper are present in the paper and/or the Supporting Information.

**Supporting Information**

The Supporting Information is available free of charge via the Internet at http://pubs.acs.org.

Sample preparation, transport measurements, NHE results along different crystal directions and at room temperature, analysis of contact misalignment and dc biased ac measurements, temperature dependences of voltage responsibility, gate dependent results for the LAO/KTO(111) Sample 2, theoretical model, calculations of carrier density, NHE for samples with different carrier densities, NHE at the $Al_2O_3$/KTO(111) interface and $SrTiO_3$(111) surface.

AUTHOR INFORMATION

**Corresponding Author**

*E-mail: rocitro@unisa.it; hepan@fudan.edu.cn; shenj5494@fudan.edu.cn

**Author Contributions**

P.H. and J.S. planed the study. J.Z. grew the films, fabricated the devices and performed transport measurements with the help from Z.Z, H.L and Y.Z.. M.T., C.A.P and R.C. performed theoretical studies. All authors discussed the results. P.H., J.F.Z, M.T., R.C., and J.S. wrote the manuscript.

**Notes**

The authors declare no competing financial interest.

†These authors contributed equally to this work.

**ACKNOWLEDGMENTS**

This work was supported by the National Key Research and Development Program of China (grants no. 2020YFA0308800 and 2022YFA1403300), National Natural Science Foundation of China (grant no.12174063), Natural Science Foundation of Shanghai (grant no. 21ZR1404300) and the start-up funding from Fudan University.